\documentclass[prl]{revtex4}
\pdfoutput=1

\usepackage{epsfig}
\usepackage{graphics}

\begin{document}
\title{Now Broadcasting in Planck Definition}
\author{Craig Hogan}
\affiliation{  University of Chicago and  Fermilab}
\begin{abstract}
If reality has finite information content, space has finite fidelity. The  quantum wave function that encodes spatial relationships may be limited to information that can be transmitted in a ``Planck  broadcast'',   with  a  bandwidth given by the inverse of the Planck time,
about $2\times 10^{43}$ bits per second.  Such a quantum system can resemble classical space-time on large scales, but locality emerges only gradually and imperfectly. Massive bodies are never perfectly  at rest, but very slightly and  slowly  fluctuate  in transverse position, with a  spectrum of variation given by the Planck time.  This distinctive new kind of  noise associated with quantum geometry would not have been noticed up to now, but may be detectable in a new kind of experiment. 
\end{abstract}
\maketitle

At the turn of the last century, Max Planck derived from first principles a universal formula for the spectrum of radiation emitted by opaque matter.  Planck's  radiation law solved a long-standing experimental mystery unexplained by classical physics, and agreed exactly with measurements.  It flowed from  a  simple,  powerful and radically new idea: that everything that happens in nature  occurs in discrete minimum packages of action, or quanta. Planck's breakthrough started the quantum revolution in physics that defined  much of twentieth century science and technology.

A few years after Planck's triumph, Albert Einstein introduced his theory of  relativity. While Planck's theory addressed the nature of matter,  Einstein's addressed  the nature of space and time.
 It  also solved  long standing mysteries, and flowed from a simple idea: that the laws of physics should not depend on how one moves.   Einstein extended his theory with another powerful idea--- that local physics is the same in any freely falling frame--- to reveal  that space and time form an active dynamical geometry, whose curvature creates the force of gravity. General Relativity was revolutionary, but it is entirely classical: Einstein's space-time is not a quantum system.
 
These two great theories of twentieth century physics have never been fully reconciled, because their core ideas are incompatible.
Relativity is based on the notion of locality, a concept not respected by quantum physics; indeed, experiments with quantum systems prove that states in reality are not localized in space. The central role of  measurement in quantum physics flies against the relativistic notion that reality is independent of an observer. 
 Perhaps most fundamentally, relativity violates quantum precepts by assigning tangible reality to unobservable things, such as events and paths in space-time.

This clash of ideas  led to agonized epistemological debates in the early part of the century, most famously between Bohr and Einstein.
 But most of physics has moved on. For  all practical purposes so far,  it works just fine to assume a continuous classical space and put quantum matter into it. That is what the well-tested Standard Model of physics does. It is a quantum theory, but only of matter, not of space-time.

Classical continuous space, as usually assumed, maps onto  real numbers: it has an infinite information density. Quantum theory suggests instead that  the information content of the world is fundamentally limited.
It is natural to suppose that all spatial relationships are just another sort of observable relationship, to be derived from the quantum theory of some system.
Let us adopt a working  hypothesis different from the usual one: {\it information in spatial position is limited by the broadcast capacity of exchanged information at a bandwidth given by a fundamental scale}. 
It is possible to work out some experimental consequences of this hypothesis even in the absence of a full theory, because the fidelity of space is limited by its information capacity.



  
  


 We have some clues to the amount of information involved. 
Planck's formula came with a new constant of nature, a fundamental unit for the quantum of action  that we now call Planck's constant, $\hbar$.
By combining his constant with Newton's constant of gravity $G$ and the speed of light $c$, Planck  obtained new ``natural'' units of length, time and mass.  In Einstein's theory, $G$ controls the dynamics of space and time.  Planck's  units therefore set the natural scale for the quantum mechanics of space-time itself,  where information about location becomes fundamentally discrete.

Because gravity is weak, the Plank scale is very small; for example,  the Planck time is $t_P= \sqrt{\hbar G/ c^5}= 5.4\times 10^{-44}$ seconds. And because that scale is so small, its quanta are very fine grained, and so far undetectable.   No experiment  shows an identifiable quantum behavior of space and time.  That is why for the last century, physicists have been able to treat space and time like a definite, continuous, classical medium. 

The lack of an experiment means that we have no guide  to   interpret mathematical ideas about blending quantum mechanics and space-time. Physicists  were  forced into the strange world of quantum mechanics   by experimental measurements, such as  radiation spectra from  black bodies  and  gases.  As Rabi said, ``Physics is an experimental science.''  So, let us ask a very practical question: 
{\it How can we build an experiment that  directly reveals the discrete character of space-time information at the Planck scale?}
The answer depends on the character of that information--- the encoding of quantum geometry in macroscopic position states.
 
Consider how  quantum systems work.
In pre-quantum physics, all properties of a system, such as positions and velocities of particles, have definite values, and change   with time according to  definite rules. In quantum physics,   as the state of a system evolves,   relationships among its properties  evolve according to  definite rules--- but in general,  individual properties  do not have definite values, even in principle. Instead, the entire system is described by a wave function of  possibilities.   Reality is that multitude of possibilities, a set of relationships.  In general,  definite, observable outcomes are impossible to predict. 




Any combined system   is literally more than the sum of its parts; a composite system contains information that  cannot be separated into information about one subsystem or another.  Information in a combined system generally resides in the correlation between its parts, a property known as ``quantum entanglement''.  
 
 The quantum challenges to conventional notions of what is real were highlighted in a famous 1935  paper by Einstein, Podolsky, and Rosen.
 Schr\"odinger  responded by introducing the idea of entanglement, as well as the provocative thought-experiment with  the uranium,  the flask of hydrocyanic acid, and the unlucky cat. As he noted, entanglement comes with nonlocality:
``Maximal knowledge of a total system does not necessarily include total knowledge of all its parts, not even when these are fully separated from each other and at the moment are not influencing each other at all.''\cite{cat}
Einstein referred to  such behavior as  ``spooky action at a distance''.



Although these ideas have created controversy over the years, it is  an experimental fact that information in the real world is not localized in space and time.
  A measurement in one place is correlated with, and affects the state of a system everywhere.\cite{horodecki,Ma2012}
There are even real-world experiments  that show examples of such quantum entanglement  between particles that never co-existed at the same time.\cite{megidish} Although experiments that demonstrate such effects are quite subtle to mount, entanglement, and the nonlocality that goes with it, are woven into the  fabric of reality. 

Einstein and  others  realized that quantum nonlocality is a big problem for relativity.\cite{wigner}  It seems to directly contradict the foundational notion of space and time, that everything happens at a definite time and place. Even the most advanced theory of space-time, General Relativity, is based on a metric that specifies the intervals between events. Quantum mechanics  implies that intervals between events, or indeed any property of events themselves,  can never be exactly measured, even in principle. And some things that happen in the world--- the quantum correlations created by interactions that we interpret as the collapse of a wave function--- are entirely delocalized in space and time.  

Physics has advanced by working around the apparent  paradox of delocalization. Often, it is not important to know exactly where something happens, since many important properties of matter do not depend on location in any particular place.
  For example, in quantum field theory,  the quantized  system is  a mode of a field wave extended in space and time--- a delocalized state. This approximation leads to extraordinarily successful  predictions  for all experiments on collisions of elementary particles at energies far below the Planck  mass, such as those at Fermilab and CERN.

Quantum delocalization  inspires  a  view of the world  made not so much of material  as of information.
This idea may be extended to space and time as well as matter.   Some properties of space and time that seem fundamental, including localization,  may actually emerge only as a macroscopic approximation,  from the flow of information in a quantum system.

Even within the  entirely classical framework of relativity and gravitation, theory has provided some hints about the possibility of such emergence, and about the significance of the Planck scale--- how  quantum mechanics blends with the physics of space and time. 
The purest states of space-time,  black holes, obey thermodynamic properties: for example, in a system  of black holes, the total area of event horizons always increases, like entropy.   More generally,  the equations of  general relativity can be derived from a purely statistical theory, by requiring  that entropy is always proportional to horizon area.\cite{Jacobson:1995ab}  Similarly, Newton's laws of motion and gravity can be derived from a  statistical theory based on entropy and coarse-graining, where information lives on surfaces and is associated with position of bodies.\cite{Verlinde:2010hp}  In these derivations, the dynamics of space-time, the  equivalence principal, and  concepts of inertia and momentum for massive bodies, all arise as emergent properties.

These results are based on essentially classical statistical arguments, but they refer to quantum information.
 The detailed quantum character of the  underlying quantum degrees of freedom associated with position in space is not known, but we can guess some of their properties.
  The precision and universality of light propagation, even across cosmic distances\cite{fermi2009}, suggests that causal structure arises from a fundamental symmetry, even if locality is only approximate.    Gravitational thermodynamics also suggest an exact number  for the  distribution and amount of information in the system: the information  fits on  two-dimensional sheets, and the total information density is the area of a sheet in Planck units.  From these arguments, we do not know how  this holographic information is encoded, but we know how much of it there is, and something about how it maps onto space.

Taken together, these theoretical ideas hint that quantum mechanics limits the amount of information in space-time.
Presumably, such a limit must  place some kind of limit on the fidelity of space-time itself:
not all classically described  locations are physically different from each other.

Video buffs know that higher bandwidth gives you a better picture. Suppose  that the bandwidth of information transmission is limited   by the Planck frequency: $\omega_P\equiv t_P^{-1} \equiv \sqrt{c^5/\hbar G}= 1.85\times 10^{43} $ Hertz. 
If this is the best that the cosmic Internet Service Provider can give us,  we do not get a perfect picture.  There is only a finite amount of information in the positional relationships of material bodies. Perhaps  a very careful experiment, that looks at position  closely in the right way,  might be able to see a bit of blurring, a lack of sharpness and clarity, like a little extra noise in the image or clipping in the cosmic sound track. The properties of the clipping may even reveal something about the compression or encoding algorithm.

An effect like that would of course be interesting to physicists, who are essentially hobbyists of nature. We used to say that physics is about discovering laws of nature, but these days we could just as well say that it is all about figuring out how the system of the universe works--- how its instructions are encoded, and what operating system it runs on. An experiment would  provide some useful clues.

Imagine then that the real world is the ultimate 4-dimensional video display. How good is it?

At first you might guess that the Planck bandwidth limit would simply create a system  with Planck size pixels everywhere; that is, a frame refresh rate given by the Planck time $t_P$, and pixel (or  voxel) size given by the Planck length, $ct_P\equiv \sqrt{\hbar G/c^3}= 1.6\times 10^{-35}$ meters, in each of the three space dimensions.  

To achieve such a fine grained picture, we  need a Planck bandwidth channel for every pixel--- a Planck density of information in four dimensions, or a Planck bandwidth in every three dimensional Planck volume. This value is the amount of information in the standard model of quantum fields, if we include all frequencies up to the Planck scale.

But this guess does not agree with  holographic emergence.
Our radically different hypothesis is that space and time  are  created from  information propagating with Planck bandwidth.  In such a   ``Planck broadcast'', space  is not assumed to exist {\it a priori}, but is a set of relationships that emerges from Planck-limited information processing.    Instead of a world   densely packed with Planck size cells, as in field theory,  perhaps  positions in space and time only contain  the amount of information that can be carried on a Planck frequency carrier wave.  In that case,  a large spatial volume has a  much smaller density of information.

Imagine  that someone sends broadcast video at the Planck frequency.  That is only enough information to refresh one pixel every Planck time.  For a larger  screen,  the refresh rate and  resolution  get worse.  How much worse?

If the broadcast encodes all of real space, it needs to encode all directions. 
Suppose that the  video screen is a sphere of  with a radius  $L$ about our broadcast point, and has pixels of size $\Delta x$.   We encode the information on the screen to refresh more slowly,  with a refresh interval given by the  time it takes light to get to the screen and back, $\tau=2L/c$---  the slowest acceptable rate for encoding a position at this distance.
Then the minimum pixel size is given by setting the total number of pixels $4\pi L^2 \Delta x^{-2}$ per time $2L/c$ equal to the Planck information rate $t_P^{-1}$,  so the pixel size is $\Delta x= \sqrt{2\pi L c t_P}$--- very small, but still much larger than the Planck length. 

Of course, nature is not really pixelated in little squares, but the same answer for the blurring scale    emerges from a more realistic  physical model   based on  waves.
Positions  encoded by wave functions that have a cutoff or bandwidth limit  convey only a limited amount of transverse spatial information  from one place to another.\cite{Hogan:2007pk,Hogan:2008zw}

Imagine a wave  that passes through a
a pair of narrow slits.   The wave  creates an interference pattern on a screen at distance $L$ that depends on  (i.e., encodes) the transverse separation of the two slits.  However, there is a resolution limit:  if the two slits have  transverse separation much smaller than  
\begin{equation}\label{varx}
\Delta x_\perp \approx  \sqrt{  L ct_P},
\end{equation}
 the interference pattern of radiation at frequency $\omega_P$ is  not distinguishable from that of a single slit.  The resolution limit from this point of view is  a diffraction limit in wave mechanics, but it is really an information bound:  the waves simply do not have enough information to resolve smaller transverse distances than that.  Notice that the distance to the screen--- and causal structure--- can be defined with much higher precision $\approx ct_P$,  by counting wave fronts. The transverse resolution, the slit separation (Eq. \ref{varx}), gets much poorer at large $L$.
  
  The corresponding angular uncertainty,
 \begin{equation}\label{vartheta}
\Delta \theta = \Delta x_\perp/ L  \approx  \sqrt{ct_P / L},
\end{equation}
gets smaller on large scales.  
Thus,  angles get {\it sharper} at larger separation, so
the notion of direction emerges more clearly on larger scales.
The total amount of angular information  grows, but only linearly with $L$,  more slowly than it would for a display with Planck size pixels.

Overall there are about $L/ct_P$ degrees of freedom corresponding to radial separation, and $L/ct_P$ corresponding to angle.
For each Planck time in duration  or radial separation, there are $L/ct_P$ angular degrees of freedom.
 The total amount of information is the number of directions times the duration, so it grows holographically, like $(L/ct_P)^2$. 
  The density of information is constant on surfaces, but in 3D space it  thins out
  with time and distance as it spreads.  
 
 This holographic scaling  is just what is needed for the statistics of emergent gravity to work.
   If we invoke that idea to set the scale of information density, the prediction for transverse mean square position uncertainty becomes very precise\cite{Hogan:2012ne}:
 \begin{equation}\label{exact}
\langle {\hat x}_\perp^2 \rangle=   L ct_P/\sqrt{4\pi}= (2.135 \times 10^{-18} {\rm m})^2 (L/{\rm 1 m}) ,
\end{equation}
with no free parameters.
We don't know the character of the actual quantum theory that controls geometry, but this estimate of the transverse blurring scale  is relatively robust, because it is just determined by the amount of information.

Apparently, if space-time is a quantum system with limited information--- a Planck broadcast---  there should be a new kind of quantum fuzziness of positions, not just for small particles, but for everything,  even for large masses. 
The blurring  is larger for larger $L$: the position resolution gets worse at larger distances.  In a laboratory size system,  it is  much larger than the Planck length---  about an attometer in scale, a billionth of a billionth of a meter.

There is vastly less  information in this macroscopic quantum system than  in standard theory--- that is, a system of quantum fields in  classical space-time with a Planck cut off--- but there is enough angular information  to agree with the apparent sharpness and classical behavior of space, as measured in  experiments to date. 
  If things could be measured at separations on  the Planck scale, the angular uncertainty would be huge; directions are not even really well defined, and it is essentially a 2D holographic system. On the scale explored by particle colliders, about $10^{16}$ times larger than Planck length, things are already very close to classical; angular blurring is too small to detect with particle experiments of limited precision, and in any case the particle masses are  small so standard quantum effects overwhelm the geometrical ones.
  
   Indeed, the new Planck blurring is {\it always} negligible compared to  standard quantum uncertainty (which does depend on mass) for systems much smaller than about a Planck mass,  $m_P=\sqrt{\hbar G/c}= 2.176\times 10^{-8}$ kg.\cite{Hogan:2012ne} In measurements  of small numbers of particles, the geometrical effects are not detectable. Unlike standard quantum effects, the Planck information limit is only important for {\it large} masses.

At first, it also seems strange that the resolution  depends on a macroscopic separation.  Intuition suggests that the state of affairs of matter and energy should not depend on how far away it is; after all, how can it ``know'' where we, the Planck broadcaster or observer, are?  According to Einstein, the laws of physics ought  to be independent of the location and motion of an observer. 

A related  worry is that  an attometer scale uncertainty, while small,  is really not all that small by the standards of particle physics.  That scale is now routinely resolved by particle colliders, like the Tevatron and the LHC.  Yet there is no sign in experiments of a new kind of fuzziness in space-time. Indeed, if we set $L$ comparable to the size the universe, we find that $\Delta x$ is actually on a  scale you can see with your own eyes, of the order of 0.1mm, the width of a hair.  Space certainly doesn't  display any  lack of sharpness on that scale when you look around.

These worries may be resolved  by invoking entanglement. Space-time is the ultimate, universal entangled system. Locality itself can emerge, via entanglement, as an approximate behavior on large scales. 

 Information is not  localized in space, but  resides in non localized correlations. 
 The density of information can depend on scale, and can be smaller for  larger systems. The effective fidelity of space-time can change depending on where something is relative to an observer.  A  measurement confined to a small volume does not know or care about a transverse geometrical displacement relative to some distant place, so the uncertainty is not observable in local measurements.  
  
 In quantum mechanics, measurements  make projections--- in Copenhagen language, they ``collapse'' the wave function. Until they are made, there is  uncertainty given by the width of the wave function--- in our case, the scale-dependent blurring. In an emergent space-time, every  world-line   defines a particular projection of the wave function associated with the  structure   of nested light cones (or ``nested causal diamonds'') around it. 

Thus, the  quantum geometrical position information is entangled for bodies whose world-lines  are close together.  If you measure the transverse position state of one massive body, you will find almost the same projection of the geometrical state for any body nearby.
That does not mean that any two bodies are in  the same position; it only means that their quantum deviation from the classical position is almost the same, relative to any arbitrary far away point.  The local relationships of the bodies in space are changed very little from standard quantum mechanics.

A classical space-time is the limiting case of a fully coherent system.  The approach to the classical limit however reveals slight departures from classical behavior that are not present in standard theory.  Nonlocal projections of the quantum state in different directions are slightly different, even on large scales.

 As the system unfolds in time, the uncertainty leads to random variations--- a new kind of noise in position measurements to distant bodies in different directions. 
 The positions of nearby bodies change together, carried along with the geometry, into the same new definite state.   Local measurements are not affected by this collective  change of position---   a new kind of  ``movement without motion''.

This interpretation of the angular uncertainty  opens up a way to build an experiment that probes Planck scale physics.
The Planck broadcast model of quantum geometry  predicts that positions  fluctuate, with a power spectrum of  angular variations given approximately  by the Planck time--- that is, in an average over duration $\tau$, the mean square variation is
\begin{equation}
\langle\Delta \theta^2\rangle_\tau\approx t_P/\tau.
\end{equation}
In an experiment of size $L$, the variations accumulate up to durations $\tau\approx L/c$, ultimately leading to variance in position given by the overall uncertainty, Eq. (\ref{exact}).
This prediction can be tested by making very sensitive measurements of  transverse positions of massive bodies. The measurement process must make a nonlocal comparison of position in different directions.

An experiment designed to detect or rule out fluctuations with these properties, called the Fermilab Holometer, is currently being developed.\cite{holometer} It uses a  technique  based on laser interferometers like those used to measure gravitational waves.  
The intensity of light emerging from a Michelson interferometer allows a precise and coherent measurement of  the positions of mirrors over an extended region of space, in this case,  40 meters in two directions.  The precision of such devices is extraordinary; they can detect variations in mean position differences on the order of attometers,  limited primarily by the quantum character of the laser light. 
 In the Holometer, correlations are measured between the signals of  two adjacent, aligned interferometers.  The correlations are sensitive to tiny, random in-common motions that change very quickly, on timescales comparable  to a light-crossing time, less than a microsecond. 
 (On longer timescales, entanglement-driven locality reduces the variation.) The  effective speed of the motionless movement  is tiny--- comparable to continental drift, only centimeters per year.

Because of quantum entanglement, the holographic noise created by the Planck broadcast information limit  creates  tiny, rapid fluctuations in  signals from the two adjacent interferometers that are coherent with each other, even if there is no connection between the devices apart from proximity.   Arguments like those outlined here, based on information in holographic emergent space, can be used to make an exact prediction for the  expected cross-correlated noise spectrum, even without knowing   details of the   fundamental theory\cite{Hogan:2012ne,Hogan:2010zs}.

Whether or not new Planck scale holographic noise is detected, the Holometer is interesting as an exploratory experiment, because it  tests the fidelity and coherence of nonlocal spatial relationships  with Planck precision for the first time. The outcome will either reveal a signature of new Planck scale physics, or experimentally prove a coherence of macroscopic space greater than what is possible with a Planck broadcast.  We don't know what we will find.

\vfil
\eject

{}

\end{document}